\documentclass[journal]{IEEEtran}
\usepackage{cite}
\ifCLASSINFOpdf
   \usepackage[pdftex]{graphicx}
   \graphicspath{{../pdf/}{../jpeg/}}
   \DeclareGraphicsExtensions{.pdf,.jpeg,.png}
\else
   \usepackage[dvips]{graphicx}
   \graphicspath{{../eps/}}
   \DeclareGraphicsExtensions{.eps}
\fi
\usepackage{enumerate}
\usepackage{amsfonts}
\usepackage{booktabs}
\usepackage[cmex10]{amsmath}

\usepackage{mathabx}
\begin{document}
\title{Energy-Efficient User Association with Open Loop Power Control for Uplink Heterogeneous Cellular Networks}

\author{Tianqing~Zhou,~
        Yongming~Huang,~\IEEEmembership{Member,~IEEE,}
        and~Luxi~Yang,~\IEEEmembership{Member,~IEEE,}% <-this % stops a space
\thanks{T. Zhou is with the School of Information Science and Engineering, Southeast University, Nanjing 210096, China, and also with the State Key Laboratory of Millimeter Waves, Department of Radio Engineering, Southeast University (e-mail:zhoutian930@163.com).}% <-this % stops a space
\thanks{Y. Huang is with the School of Information Science and Engineering, Southeast University, Nanjing 210096, China, and also with the Key Laboratory of BroadbandWireless Communication and Sensor Network Technology (Nanjing University of Posts and Telecommunications), Ministry of Education (e-mail:huangym@seu.edu.cn)}% <-this % stops a space
\thanks{L.Yang is with the School of Information Science and Engineering, Southeast University, Nanjing 210096, China (e-mail:lxyang@seu.edu.cn).}}

\maketitle

\begin{abstract}
Energy reduction for wireless systems becomes more and more important due to its impact on the operation cost and global carbon footprint. In this paper, we investigate three kinds of energy-efficient association schemes under open loop power control for uplink heterogeneous cellular networks, which are formulated as a whole energy efficiency maximization problem, a sum energy efficiency maximization problem and a utility maximization problem respectively. The third case takes account of load balancing level and user fairness in the energy-efficient association. Considering that the first problem is in a fractional mixed-integer form, we introduce an energy efficiency parameter to convert it into a parametric subtractive form, and then design an effective iterative algorithm to achieve the optimal solutions. As for the third problem, we first introduce a dual variable to decouple the constraint and then develop a distributed algorithm using dual decomposition. In addition, we also give the convergence proofs for the proposed algorithms. In order to confirm the effectiveness of energy-efficient user association algorithms, we introduce other association rules for comparison, and investigate the influences of different parameters on the association performance of these association rules.
\end{abstract}


% \fi
\IEEEpeerreviewmaketitle

\section{Introduction}
\IEEEPARstart{T}{o} meet the explosive growth of mobile data volume driven by various applications such as smartphones and tablets, the conventional wireless network based on macro base stations (BSs) has shifted to the one mixed with various low-power BSs such as pico BSs and femto BSs \cite{1,2}. These BSs differ primarily in transmit power, deployment cost, propagation characteristic and backhaul \cite{3,4}.
\par
Although the deployment of low-power BSs can improve the system capacity and user experience, it requires operators to solve some novel problems such as the design of network architecture \cite{5}, interference management \cite{6} and synchronization \cite{7}. In fact, the interference management in HCNs is one of the most challenging issues. To enhance network capacity and support user's quality-of-service (QoS) in different network tiers, the interference problem in HCNs needs to be solved. In the co-channel deployment, there exists inter-tier interference among different network tiers and intra-tier interference within each network tier (e.g., macro and pico tiers), and these interferences should be properly managed. For code-division-multiple-access (CDMA) networks \cite{8,9,10,11}, power control and user association algorithms provide efficient measures to mitigate the interference and increase system capacity.
\par
A good user association algorithm should provide an efficient mechanism to select one or several BSs for some user so that certain performance metrics (e.g., achievable rate, transmit power, energy efficiency, cell load, and geographical location) of interest are optimized. As a key metric in green communications, the energy efficiency optimization has attracted more and more mentions. During user association, the consideration of power control and/or beamforming is often beneficial to mitigating the interference and reducing the energy consumption.
\par
With the rapid development of information and communication technologies, the energy consumption is growing at staggering rate \cite{12,13,14,15,16}. The exploding growth of energy consumption in wireless communications will be potentially harmful to environment and increase operation costs. To this end, the designed association algorithms should be energy-efficient \cite{17}.
In order to reduce the energy consumption and thus realize the goal of green communications, many schemes introduce power control and/or beamforming into the association procedure. So far, the existing works on energy-efficient user association can be broadly classified into four types. Next, we will give the detailed discussions for various energy-efficient association schemes.
\par
In the first type, many schemes are designed to minimize the sum power during user association. In \cite{9}, authors jointly perform user association and power control to minimize the sum power under user's rate constraint. In addition, authors in \cite{18} jointly consider user association, power control and bandwidth allocation to minimize the sum power for uplink heterogeneous cloud wireless access networks. Instead of direct power control, authors in \cite{19} jointly consider  beamforming and user association to minimize downlink and uplink energy consumption.
\par
In the second type, designers perform user association to maximize the whole energy efficiency. The whole energy efficiency is denoted as the ratio of sum rate to total power consumption. In \cite{20}, authors design an iterative association algorithm to maximize the whole energy efficiency for uplink HCNs, which jointly considers subband selection and power control during user association.
\par
In the third type, the researchers try to design some association schemes to maximize the sum energy efficiency, where the energy efficiency represents the ratio of user's effective rate to the power consumption of some BS selected by this user. In \cite{21}, authors design a energy-efficient association scheme without power control for downlink HCNs. In this scheme, authors maximize the network-wide utility that is a logarithmic function of energy efficiency, and develop a distributed algorithm using dual decomposition. Evidently, the introduction of logarithmic utility function can guarantee the user fairness.
\par
In the last type, some schemes are advocated to optimize some hybrid targets. In \cite{22}, authors try to design an association scheme to trade off system load and power consumption, which maximizes the network load while minimizing the power consumption. In \cite{23}, authors jointly optimize network load and resource consumption, and thus design an association strategy that maximizes the network load and meanwhile minimizing the resource consumption. Based on work \cite{23}, authors in \cite{24} make a further attempt for minimizing power consumption. Instead of power control, authors in \cite{25} reduce the energy consumption through BS operation (on/off), and design an association scheme that trades off energy and delay.
\par
As revealed in \cite{26}, the schemes with joint user association and beamforming may be unreasonable since the user association often takes place at a fairly long time scale but the beamforming takes place in a shorter time scale. Evidently, the former utilizes the slow-fading channel, but the latter adopts fast-fading channel.
\par
In this paper, we jointly perform user association and power control for uplink HCNs. Contrary to the existing schemes, we adopt an open loop power control rule based on pathloss. This operation greatly reduces the complexity of association algorithm, and is beneficial to practical implement. When (direct) power control and/or beamforming are/is considered in the association problem, the design of association algorithm for achieving the optimal solutions is challenging, and the designed algorithm often has relatively high computation complexity.
\par
Under open loop power control, we investigate three types of energy-efficient association schemes that maximize the whole energy efficiency, the sum energy efficiency and the network-wide utility that is a logarithmic function of user's energy efficiency respectively. Unlike other two cases, the third case jointly considers the load balancing levels of BSs and the user fairness. The reason for this consideration is that energy-efficient association schemes enhance the impact of pathloss. According to the parameters mentioned in the numerical simulation, we can deduce that the value range of PBS's pathloss often contains the one of MBS's pathloss. To improve the achievable rate and meanwhile reducing the transmit power, the enhanced impact of pathloss leads to that more users will be associated with PBSs.
\par
Considering that the first problem is in a fractional mixed-integer form and hard to tackle, we introduce an energy efficiency parameter to convert it into a parametric subtractive form, and then design an effective algorithm to achieve the optimal solutions. As for the third problem, we first introduce a dual variable to decouple the constraint and then develop a distributed algorithm using dual decomposition. Compared with other problems, the second problem can be easily solved without any iteration. In addition, we also give the convergence proofs for the proposed algorithms. To highlight the effectiveness of energy-efficient association schemes, we introduce other association rules for comparison, which include maximal achievable rate association (MARA) and association with user fairness (AUF) \cite{27}. In the AUF, the utility is denoted as a logarithmic function with respect to effective rate. At last, we give numerical simulation and investigate the influences of different parameters on the association performance of these association rules.
\par
The remainder of this paper is organized as follows. In section II, we present the system model. In section III, we formulate the energy-efficient association problem that maximizes the whole energy efficiency, design an effective algorithm and give the convergence proof for it. In section IV, we formulate another energy-efficient association problem that maximizes the sum energy efficiency, and give the detailed association procedure. In section V, we formulate the energy-efficient association problem that maximizes the sum utility of user's energy efficiency, design a distributed algorithm and provide the convergence proof for it. In section VI, we give numerical results for different network parameters. In section VII, we present further discussions and conclusions.
\section{System Model}
\begin{figure}[!t]
\centering
\centerline{\includegraphics[width=4in]{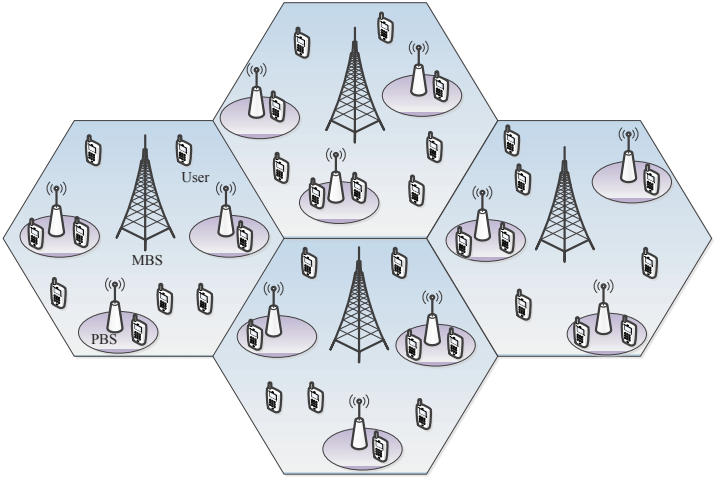}}
\caption{Illustration of two-tier heterogeneous cellular networks.}
\label{fig1}
\end{figure}

In this paper, we consider two-tier HCNs consisting of macro BSs (MBSs) and pico BSs (PBSs), which is illustrated in Fig. \ref{fig1}. In such HCNs, the MBSs are deployed according to a conventional cellular framework, while the users and PBSs are scattered into each macrocell in a relatively random manner. So far, many efforts in the literature consider an irregular deployment obtained by a random spatial point process. Such deployment is beneficial to the derivation of tractable expressions, and thus has been widely used for performance analyses. In fact, regular deployment and irregular deployment have not any impact on the essence of association scheme.
\par
We let $\mathcal{N}$ be the set of BSs including MBSs and PBSs. In addition, the set of users is $\mathcal{K}$, the cardinalities of $\mathcal{N}$ and $\mathcal{K}$ are denoted as $N=\left| \mathcal{N} \right|$ and $K=\left| \mathcal{K} \right|$ respectively. Thus, the signal-interference-and-noise (SINR) of user $k$ at BS $n$ can be written as
\begin{equation}\label{eq1}
{{\text{SINR}}_{nk}}=\frac{\kappa {{p}_{nk}}{{g}_{nk}}}{{{\sum }_{j\in \mathcal{K}\backslash \left\{ k \right\}}}{{p}_{nj}}{{g}_{nj}}+\sigma _{n}^{2}},
\end{equation}
where $\kappa $ is the processing gain that represents the ratio of the spreading bandwidth to the symbol rate \cite{28}; ${{p}_{nk}}=\min \left\{ {{\Gamma }_{k}}*{{\sigma }_{n}}^{2}/{{10}^{-{{l}_{nk}}/10}},{{10}^{p_{k}^{\max}/10}}\  \right\}$ \cite{24} represents the desired transmit power of user $k$ when it is associated with BS $n$, the target signal-to-noise ratio (SNR) ${{\Gamma }_{k}}$ of each user is set to be 10 dB, ${{l}_{nk}}$ represents the pathloss in dB between user $k$ and BS $n$, and the maximal transmit power ${p_{k}^{\max}}$ of user $k$ is set to be 23 dBm; ${{g}_{nk}}$ denotes the channel gain between user $k$ and BS $n$; $\sigma _{n}^{2}$ is the noise power of BS $n$. Then, the achievable rate of user $k$ at BS $n$ is
\begin{equation}\label{eq2}
{{r}_{nk}}={{\log }_{2}}\left( 1+\text{SIN}{{\text{R}}_{nk}} \right).
\end{equation}
\par
Next, we will study three types of energy-efficient association schemes under open loop power control for uplink HCNs. These schemes include association with maximizing the whole energy efficiency (AMWEE), association with maximizing the sum energy efficiency (AMSEE) and energy-efficient association with user fairness (EEAUF). In the scheme EEAUF, we maximize the sum utility that is the logarithmic function of user's energy efficiency. As for these three schemes, we design detailed algorithms and give the convergence proofs of algorithms.
\section{Association with Maximizing the Whole Energy Efficiency}
In this section, we will design an energy-efficient association scheme to maximize the whole energy efficiency under open loop power control, where the whole energy efficiency is denoted as the ratio of the sum of achievable rates to the total power consumption. Mathematically, this scheme is formulated as
\begin{equation}\label{eq3}
\begin{split}
  \underset{\boldsymbol{x}}{\mathop{\max }}\,&\ E\left( \boldsymbol{x} \right)=\frac{{{E}_{1}}\left( \boldsymbol{x} \right)}{{{E}_{2}}\left( \boldsymbol{x} \right)}=\frac{\sum\limits_{n\in \mathcal{N}}{\sum\limits_{k\in \mathcal{K}}{{{x}_{nk}}{{r}_{nk}}}}}{\sum\limits_{n\in \mathcal{N}}{\sum\limits_{k\in \mathcal{K}}{{{x}_{nk}}{{p}_{nk}}}}+K{{p}_{c}}} \\
\text{s.t.}\ & \sum\limits_{n\in \mathcal{N}}{{{x}_{nk}}}=1,\forall k\in \mathcal{K}, \\
 & {{x}_{nk}}\in \left\{ 0,1 \right\},\forall n\in \mathcal{N},\forall k\in \mathcal{K}, \\
\end{split}
\end{equation}
where the $E$ represents the whole energy efficiency; $\boldsymbol{x}=\left\{ {{x}_{nk}},\forall n\in \mathcal{N},\forall k\in \mathcal{K} \right\}$; ${{x}_{nk}}$ is an association indicator, i.e., ${{x}_{nk}}=1$ when user $k$ is associated with BS $n$, 0 otherwise; the first constraint shows that one user can just be associated with one BS.
\par
Through direct observation, it is easy to find that the formulated problem is in a nonlinear and mixed-integer form, thus this problem is non-convex and hard to tackle. According to the form of the formulated problem, we can conclude that it belongs to a classical fractional programming problem. Thus, we can adopt a common approach \cite{29,30,31} to solve it, which transforms the original problem into a linear programming through nonlinear variable transformation. By exploiting the relationship between parametric programming problem and fractional programming problem, we reformulate the problem \eqref{eq3} into the following univariate equation:
\begin{equation}\label{eq4}
F\left( \gamma  \right)=\underset{\boldsymbol{x}\in \mathbb{D}}{\mathop{\max }}\,\ \left\{ {{E}_{1}}\left(\boldsymbol{x} \right)-\gamma {{E}_{2}}\left(\boldsymbol{x} \right) \right\}=0,
\end{equation}
where $\mathbb{D}$ is the feasible domain of problem \eqref{eq3}. The equivalence between problem \eqref{eq3} and problem \eqref{eq4} is given by the following Theorem 1.
\\
\par
\noindent
\textbf{Theorem 1.} The following two statements are equivalent:
\par
\noindent
(a) $\underset{\boldsymbol{x}\in \mathbb{D}}{\mathop{\max }}\,\ E\left(\boldsymbol{x} \right)=\underset{\boldsymbol{x}\in \mathbb{D}}{\mathop{\max }}\,\ \frac{{{E}_{1}}\left(\boldsymbol{x} \right)}{{{E}_{2}}\left(\boldsymbol{x} \right)}=\gamma$.
\par
\noindent
(b) $F\left( \gamma  \right)=\underset{\boldsymbol{x}\in \mathbb{D}}{\mathop{\max }}\,\ \left\{ {{E}_{1}}\left(\boldsymbol{x} \right)-\gamma {{E}_{2}}\left(\boldsymbol{x} \right) \right\}=0$.
\par
\emph{Proof: }  We first prove the sufficient condition of Theorem 1, i.e., (b) can be deduced from (a). Let ${{\boldsymbol{x}}^{*}}$ be the solution of the problem \eqref{eq3}, we have the following result for all $\boldsymbol{x}$ in feasible domain $\mathbb{D}$, which is given by
\begin{equation}\label{eq5}
\gamma =F\left({{\boldsymbol{x}}^{*}} \right)=\frac{{{E}_{1}}\left({{\boldsymbol{x}}^{*}} \right)}{{{E}_{2}}\left({{\boldsymbol{x}}^{*}} \right)}\ge \frac{{{E}_{1}}\left(\boldsymbol{x} \right)}{{{E}_{2}}\left(\boldsymbol{x} \right)}.
\end{equation}
\par
Since the total power consumption is greater than zero, we can easily find that ${{E}_{2}}\left(\boldsymbol{x} \right)>0$ for all $\boldsymbol{x}$ in feasible domain $\mathbb{D}$. Thus, we have
\begin{equation}\label{eq6}
\left\{ \begin{split}
  & {{E}_{1}}\left( {{\boldsymbol{x}}^{*}} \right)-\gamma {{E}_{2}}\left( {{\boldsymbol{x}}^{*}} \right)=0 \\
 & {{E}_{1}}\left( \boldsymbol{x} \right)-\gamma {{E}_{2}}\left( \boldsymbol{x} \right)\le 0. \\
\end{split} \right.
\end{equation}
\par
According to the equation \eqref{eq6}, we can easily conclude that $F\left( \gamma  \right)=\underset{\boldsymbol{x}\in \mathbb{D}}{\mathop{\max }}\,\ \left\{ {{E}_{1}}\left(\boldsymbol{x} \right)-\gamma {{E}_{2}}\left(\boldsymbol{x} \right) \right\}=0$ and the maximal value is achieved at ${{\boldsymbol{x}}^{*}}$.
\par
Next, we prove the necessary condition of Theorem 1, i.e., (a) can be deduced from (b). Let ${{\boldsymbol{x}}^{*}}$ be the solutions of problem \eqref{eq4}, we have the following result for all $\boldsymbol{x}$ in feasible domain $\mathbb{D}$, which is presented as
\begin{equation}\label{eq7}
\begin{split}
  0=F\left( \gamma  \right)&= {{E}_{1}}\left( {{\boldsymbol{x}}^{*}} \right)-\gamma {{E}_{2}}\left( {{\boldsymbol{x}}^{*}} \right) \\
 &\ge {{E}_{1}}\left( \boldsymbol{x} \right)-\gamma {{E}_{2}}\left( \boldsymbol{x} \right). \\
\end{split}
\end{equation}
\par
Then, we have
\begin{equation}\label{eq8}
\frac{{{E}_{1}}\left({{\boldsymbol{x}}^{*}} \right)}{{{E}_{2}}\left({{\boldsymbol{x}}^{*}} \right)}=\gamma \text{ and }\frac{{{E}_{1}}\left(\boldsymbol{x} \right)}{{{E}_{2}}\left(\boldsymbol{x} \right)}\le \gamma .
\end{equation}
\par
Based on the equation \eqref{eq8}, we can easily know that $\underset{\boldsymbol{x}\in \mathbb{D}}{\mathop{\max }}\,\ \frac{{{E}_{1}}\left(\boldsymbol{x} \right)}{{{E}_{2}}\left(\boldsymbol{x} \right)}=\gamma$ and the maximal value is achieved at ${{\boldsymbol{x}}^{*}}$.
\par
The above-mentioned theorem means that the univariate equation $F\left( \gamma  \right)=0$ is essentially equivalent to the original problem \eqref{eq3}. In other words, if we can find a parameter $\gamma$ that satisfies $F\left( \gamma  \right)=0$, then the optimal solution of problem \eqref{eq4} is also the optimal one of problem \eqref{eq3}. Based on the optimal condition mentioned in Theorem 1, the problem \eqref{eq3} can be equivalent to
\begin{equation}\label{eq9}
\begin{split}
  \underset{\boldsymbol{x},\gamma }{\mathop{\max }}\,&\ {{E}_{1}}\left( \boldsymbol{x} \right)-\gamma {{E}_{2}}\left( \boldsymbol{x} \right) \\
\text{s.t.} &\ \sum\limits_{n\in \mathcal{N}}{{{x}_{nk}}}=1,\forall k\in \mathcal{K}, \\
 & {{x}_{nk}}\in \left\{ 0,1 \right\},\forall n\in \mathcal{N},\forall k\in \mathcal{K}, \\
\end{split}
\end{equation}
It is worthwhile to note that the problems \eqref{eq4} and \eqref{eq9} are equivalent. As for this problem, some results can be found in the following lemma.
\\
\par
\noindent
\textbf{Lemma 1.} For all feasible $\boldsymbol{x}$ and $\gamma $, $F\left( \gamma  \right)$ is a strictly decreasing function with respect to $\gamma $ and $F\left( \gamma  \right)\ge 0$.
\par
\emph{Proof: }  We assume that ${{\gamma }_{1}}$ and ${{\gamma }_{2}}$ are the optimal solution for the optimal associations ${{\boldsymbol{x}}^{\dagger }}$ and ${{\boldsymbol{x}}^{*}}$, where ${{\gamma }_{1}}>{{\gamma }_{2}}$. Then, we have
\begin{equation}\label{eq10}
\begin{split}
  F\left( {{\gamma }_{2}} \right)&={{E}_{1}}\left( {{\boldsymbol{x}}^{*}} \right)-{{\gamma }_{2}}{{E}_{2}}\left( {{\boldsymbol{x}}^{*}} \right) \\
 &>{{E}_{1}}\left( {{\boldsymbol{x}}^{\dagger }} \right)-{{\gamma }_{2}}{{E}_{2}}\left( {{\boldsymbol{x}}^{\dagger }} \right) \\
 &>{{E}_{1}}\left( {{\boldsymbol{x}}^{\dagger }} \right)-{{\gamma }_{1}}{{E}_{2}}\left( {{\boldsymbol{x}}^{\dagger }} \right) \\
 &=F\left( {{\gamma }_{1}} \right) \\
\end{split}
\end{equation}
\par
Thus, $F\left( \gamma  \right)$ is a strictly decreasing function with respect to $\gamma $. In addition, we let $\boldsymbol{\tilde{x}}$ be any association and set $\tilde{\gamma }=\frac{{{E}_{1}}\left( {\boldsymbol{\tilde{x}}} \right)}{{{E}_{2}}\left( {\boldsymbol{\tilde{x}}} \right)}$, then
\begin{equation}\label{eq11}
\begin{split}
  F\left( {\tilde{\gamma }} \right)&=\underset{\boldsymbol{x}\in \mathbb{D}}{\mathop{\max }}\,{{E}_{1}}\left( \boldsymbol{x} \right)-\tilde{\gamma }{{E}_{2}}\left( \boldsymbol{x} \right) \\
 &\ge {{E}_{1}}\left( {\boldsymbol{\tilde{x}}} \right)-\tilde{\gamma }{{E}_{2}}\left( {\boldsymbol{\tilde{x}}} \right) \\
 &=0 \\
\end{split}
\end{equation}
\par
Thus, we have $F\left( \gamma  \right)\ge 0$.
\par
Due to the constraints of problem \eqref{eq3}, the feasible domain of $\boldsymbol{x}$ should be a discrete and finite set consisting of all possible associations, and thus $F\left( \gamma  \right)$ is a continuous but non-differentiable function in terms of $\gamma $.
\par
Seen from the problem \eqref{eq9}, the optimal ${{\gamma }^{*}}$ cannot be achieved directly. Although the optimal ${{\gamma }^{*}}$ cannot be obtained in a closed-form expression, there exist two types of methods to find the feasible solution. In the first type, we can use bisection method to search the solution in feasible domain \cite{32}. In the other type, we can adopt an iterative algorithm to update $\gamma $ while ensuring that the corresponding solution $\boldsymbol{x}$ remains feasible in each iteration \cite{33}. Considering that it is difficult for designers to find the tight bound of $\gamma $ in the first type, we primarily focus on the second type.
\par
When the optimal energy efficiency parameter $\gamma $ is given, the problem \eqref{eq9} can be simplified into
\begin{equation}\label{eq12}
\begin{split}
  \underset{\boldsymbol{x}}{\mathop{\max }}\,& \ \sum\limits_{n\in \mathcal{N}}{\sum\limits_{k\in \mathcal{K}}{{{x}_{nk}}{{\hbar }_{nk}}}} \\
 \text{s.t.}&\ \sum\limits_{n\in \mathcal{N}}{{{x}_{nk}}}=1,\forall k\in \mathcal{K}, \\
 & {{x}_{nk}}\in \left\{ 0,1 \right\},\forall n\in \mathcal{N},\forall k\in \mathcal{K}, \\
\end{split}
\end{equation}
where ${{\hbar }_{nk}}={{r}_{nk}}-\gamma {{p}_{nk}}$.
\par
Thus, the optimal association indicator ${{\boldsymbol{x}}^{*}}$ can be expressed as
\begin{equation}\label{eq13}
x_{nk}^{*}=\left\{ \begin{split}
  & 1,n=\arg \underset{n\in \mathcal{N}}{\mathop{\max }}\,{{\hbar }_{nk}},k\in \mathcal{K}, \\
 & 0,\text{otherwise}, \\
\end{split} \right.
\end{equation}
The rule \eqref{eq13} shows that any user $k$ selects some BS $n$ to maximize the obtained utility ${{\hbar }_{nk}}$. Once the optimal association indicator is found, we can update $\gamma $ using the following rule:
\begin{equation}\label{eq14}
{{\gamma }^{t+1}}=\frac{{{E}_{1}}\left( {{\boldsymbol{x}}^{t}} \right)}{{{E}_{2}}\left( {{\boldsymbol{x}}^{t}} \right)}.
\end{equation}
\par
The detailed procedure for updating energy efficiency parameter and association indicator can be found in Algorithm 1, where $t$ is iteration index; ${{T}_{1}}$ is the maximal number of iterations.
\begin{table}[h]
\centering
%\caption{Algorithm 1: AMWEE}
\begin{tabular}{ll}
\toprule[1pt]
\textbf{Algorithm 1} AMWEE\\ \midrule[0.5pt]
1: \textbf{Initialization:} $t=0$, ${{\gamma }^{t}}$, ${{T}_{1}}$. \\
2: \textbf{Repeat(Main Loop)}\\
3:\ \ \ \ Set $\boldsymbol{x}=\boldsymbol{0}$. \\
4:\ \ \ \ \textbf{For} $k\in \mathcal{K}$ \\
5:\ \ \ \ \ \ \ \ User $k$ selects BS ${{n}^{*}}$ that satisfies the following condition:\\
6:\ \ \ \ \ \ \ \ \ \ \ \ \ ${{n}^{*}}=\arg \underset{n\in \mathcal{N}}{\mathop{\max }}\,{{\hbar }_{nk}}$.\\
7:\ \ \ \ \ \ \ \ Set ${{x}_{{{n}^{*}}k}}=1$.\\
8:\ \ \ \ \textbf{EndFor} \\
9:\ \ \ \ Set ${{\gamma }^{t+1}}=\frac{{{E}_{1}}\left( {{\boldsymbol{x}}^{t}} \right)}{{{E}_{2}}\left( {{\boldsymbol{x}}^{t}} \right)}$.\\
10:\ \ \ \ $t=t+1$ \\
11:\textbf{Until} convergence or $t={{T}_{1}}$\\ \bottomrule[0.5pt]
\end{tabular}
\label{tab1}
\end{table}
\par
Next, we will give the convergence proof of Algorithm 1. To this end, we first provide the following results.
\\
\par
\noindent
\textbf{Theorem 2.} Let ${{\boldsymbol{x}}^{t}}$ be the maximizer of the problem \eqref{eq9} for the energy efficiency parameter ${{\gamma }^{t}}$ at the t-th iteration. If this parameter at the (t+1)-th iteration is updated by the rule \eqref{eq14}, then ${{\gamma }^{t}}$ is monotonically increasing with respect to $t$.
\par
\emph{Proof: } We assume that ${{\boldsymbol{x}}^{t}}$ is the optimal association at t-th iteration. According to the Theorem 1, we have
\begin{equation}\label{eq15}
F\left( {{\gamma }^{t}} \right)={{E}_{1}}\left( {{\boldsymbol{x}}^{t}} \right)-{{\gamma }^{t}}{{E}_{2}}\left( {{\boldsymbol{x}}^{t}} \right).
\end{equation}
\par
By employing the formula \eqref{eq14}, the equation \eqref{eq15} can be converted into
\begin{equation}\label{eq16}
F\left( {{\gamma }^{t}} \right)=\left( {{\gamma }^{t+1}}-{{\gamma }^{t}} \right){{E}_{2}}\left( {{\boldsymbol{x}}^{t}} \right).
\end{equation}
\par
Since ${{E}_{2}}\left(\boldsymbol{x} \right)>0$ and the Lemma 1 has shown that $F\left( {{\gamma }^{t}} \right)\ge 0$, we have ${{\gamma }^{t+1}}\ge {{\gamma }^{t}}$.
\\
\par
\noindent
\textbf{Lemma 2.} Under the rule \eqref{eq14}, the Algorithm 1 finally converges to a stable point.
\par
\emph{Proof: }According to the Theorem 2, we know that ${{\gamma }^{t}}$ is monotonically increasing with respect to $t$. Moreover, ${{E}_{1}}$ and ${{E}_{2}}$ should be bounded in real system, and thus the $\gamma $ is also bounded. According to the monotonic boundary sequence theorem \cite{34}, it is easy to find that the Algorithm 1 is convergent.
\section{Association with Maximizing Sum Energy Efficiency}
In this section, we will design an energy-efficient association scheme to maximize the sum energy efficiency under open loop power control, where the (individual) energy efficiency of some user is denoted as the ratio of its achievable rate to its power consumption. Mathematically, this scheme is formulated as
\begin{equation}\label{eq17}
\begin{split}
 \underset{\boldsymbol{x}}{\mathop{\max }}\,& \sum\limits_{n\in \mathcal{N}}{\sum\limits_{k\in \mathcal{K}}{{{x}_{nk}}{{\Xi}_{nk}}}}=\sum\limits_{n\in \mathcal{N}}{\sum\limits_{k\in \mathcal{K}}{{{x}_{nk}}\frac{{{r}_{nk}}}{{{p}_{nk}}+{{p}_{c}}}}} \\
\text{s.t.}&\sum\limits_{n\in \mathcal{N}}{{{x}_{nk}}}=1,\ \ \forall k\in \mathcal{K}, \\
 &{{x}_{nk}}\in \left\{ 0,1 \right\},\ \ \forall n\in \mathcal{N},\forall k\in \mathcal{K}, \\
\end{split}
\end{equation}
where ${{\Xi}_{nk}}$ is the energy efficiency of user $k$ at BS $n$; the energy efficiency of user $k$ is $\sum\nolimits_{n\in \mathcal{N}}{{{x}_{nk}}{{\Xi}_{nk}}}$.
\par
Evidently, the optimal association indicator ${{\boldsymbol{x}}^{*}}$ can be expressed as
\begin{equation}\label{eq18}
x_{nk}^{*}=\left\{ \begin{split}
  & 1,n=\arg \underset{n\in \mathcal{N}}{\mathop{\max }}\,{{\Xi}_{nk}},k\in \mathcal{K}, \\
 & 0,\text{otherwise}, \\
\end{split} \right.
\end{equation}
\par
The rule \eqref{eq18} shows that any user $k$ selects some BS $n$ to maximize the obtained energy efficiency ${{E}_{nk}}$. The detailed association can be found in Algorithm 2. Note that the Algorithm 2 is not iterative algorithm and is very easy to be implemented in real system.
\begin{table}[h]
\centering
%\caption{Algorithm 2: AMSEE}
\begin{tabular}{ll}
\toprule[1pt]
\textbf{Algorithm 2} AMSEE\\ \midrule[0.5pt]
1: \textbf{Initialization:} $\boldsymbol{x}=\boldsymbol{0}$. \\
2:\ \textbf{For} $k\in \mathcal{K}$ \\
3:\ \ \ \ User $k$ selects BS ${{n}^{*}}$ that satisfies the following condition:\\
4:\ \ \ \ \ \ \ \ ${{n}^{*}}=\arg \underset{n\in \mathcal{N}}{\mathop{\max }}\,{{\Xi}_{nk}}.$\\
5:\ \ \ \ Set ${{x}_{{{n}^{*}}k}}=1$.\\
6:\ \textbf{EndFor} \\ \bottomrule[0.5pt]
\end{tabular}
\label{tab2}
\end{table}
\section{Energy-Efficient Association with User Fairness}
Considering the imbalanced load distribution in HCNs, we design an energy-efficient user association scheme for load balancing. To this end, this scheme should be related to the cell load (the number of users associated with some cell). Instead of achievable rate involved in energy efficiency expression, the effective rate is a good option for load balancing. At last, this scheme is formulated as a network-wide utility maximization problem. Mathematically, it is given by
\begin{equation}\label{eq19}
\begin{split}
  \underset{\boldsymbol{x},\boldsymbol{y}}{\mathop{\max }}\,& \ \sum\limits_{n\in \mathcal{N}}{\sum\limits_{k\in \mathcal{K}}{{{x}_{nk}}{{U}_{nk}}}}\left( \frac{{{R}_{nk}}}{{{p}_{nk}}+{{p}_{c}}} \right) \\
\text{s.t.}&\sum\limits_{n\in \mathcal{N}}{{{x}_{nk}}}=1,\ \ \forall k\in \mathcal{K}, \\
 &\sum\limits_{k\in \mathcal{K}}{{{x}_{nk}}}={{y}_{n}},\ \ \forall n\in \mathcal{N}, \\
 &{{x}_{nk}}\in \left\{ 0,1 \right\},\ \ \forall n\in \mathcal{N},\forall k\in \mathcal{K}, \\
\end{split}
\end{equation}
where ${{R}_{nk}}=\frac{{{r}_{nk}}}{{{y}_{n}}}$ represents the effective rate of user $k$ at BS $n$; $\boldsymbol{y}=\left\{ {{y}_{n}},\forall n\in \mathcal{N}\right\}$; ${{y}_{n}}$ is the number of users associated with BS $n$, i.e., load of BS $n$. \par
According to the definition of effective rate, it is easy to find that not all users can always select some BSs with the best channel condition (highest achievable rate) because of the load limit. In other words, the user's (effective) rate will decrease when the associated BS is overloaed. That's because the amount of available resources will decrease when too many users are associated with some BS. To mitigate the decrement of user's rate, some users should be offloaded to underloaded BSs from the overloaded BSs. Evidently, the proposed scheme can balance the loads among BSs, especially for various BSs.
\par
In order to guarantee the user fairness and the design of effective algorithm, we take a logarithmic function as the utility function mentioned in the formulated problem. Then, the problem \eqref{eq19} can be converted into
\begin{equation}\label{eq20}
\begin{split}
 \underset{\boldsymbol{x},\boldsymbol{y}}{\mathop{\max }}\,& \ G\left( \boldsymbol{x},\boldsymbol{y} \right)=\sum\limits_{n\in \mathcal{N}}{\sum\limits_{k\in \mathcal{K}}{{{x}_{nk}}{{h}_{nk}}}}-\sum\limits_{n\in \mathcal{N}}{{{y}_{n}}\log {{y}_{n}}} \\
\text{s.t.}&\sum\limits_{n\in \mathcal{N}}{{{x}_{nk}}}=1,\ \ \forall k\in \mathcal{K} \\
 &\sum\limits_{k\in \mathcal{K}}{{{x}_{nk}}={{y}_{n}}},\ \ \forall n\in \mathcal{N} \\
 &{{x}_{nk}}\in \left\{ 0,1 \right\},\ \ \forall n\in \mathcal{N},\forall k\in \mathcal{K} \\
\end{split}
\end{equation}
where ${{h}_{nk}}=\log {{r}_{nk}}-\log \left( {{p}_{nk}}+{{p}_{c}} \right)$.
\par
Seen from the objective function of problem \eqref{eq20}, we know this association rule trades off the system throughput and the system load. Specially, in order to achieve a high utility, any user needs to jointly consider achievable rate and load in the association. To increase the utility, users may be offloaded to other underloaded BSs with relatively high achievable rates. Unlike the association with maximal achievable rate (AMAR), the proposed association rule cannot always let users select some BSs with highest achievable rate.
\par
Next, we will give the detailed discussions on how to solve the problem \eqref{eq20}. As a general approach to solve the optimization problems, especially for large-scale problems \cite{35}, the dual decomposition method has been widely applied in resource allocation \cite{36} and transceiver design \cite{37} and routing \cite{38}. By breaking the original problem up into smaller subproblems that can often be tackled in a distributed manner, an effective algorithm can be easily designed for distributed implement. Recently, the dual decomposition method has attracted more and more attentions for user association in the literature \cite{21,26,27}, which can effectively deal with the large-scale mixed-integer optimization problems. In this section, we will utilize the dual decomposition method to design a distributed algorithm. Specially, the original problem with a complex form is cut into two subproblems that can be solved by users and BSs separately.
\par
Through direct observation, it is easy to find that the second constraint of problem \eqref{eq20} is coupling. To decouple it, we introduce a dual variable $\boldsymbol{\mu }=\{{{\mu }_{n}},n\in \mathcal{N}\}$. Then, the Lagrange function in terms of this constraint is
\begin{equation}\label{eq21}
\begin{split}
 \mathcal{L}\left( \boldsymbol{x},\boldsymbol{y},\boldsymbol{\mu } \right)=& \sum\limits_{n\in \mathcal{N}}{\sum\limits_{k\in \mathcal{K}}{{{x}_{nk}}{{h}_{nk}}}}-\sum\limits_{n\in \mathcal{N}}{{{y}_{n}}\log {{y}_{n}}} \\
 &+\sum\limits_{n\in \mathcal{N}}{{{\mu }_{n}}\left( {{y}_{n}}-\sum\limits_{k\in \mathcal{K}}{{{x}_{nk}}} \right)}, \\
\end{split}
\end{equation}
\par
Thus, the dual function can be given by
\begin{equation}\label{eq22}
H\left( \boldsymbol{\mu } \right)=\ \left\{ \begin{split}
  \underset{\boldsymbol{x},\boldsymbol{y}}{\mathop{\max }}\,&\mathcal{L}\left( \boldsymbol{x},\boldsymbol{y},\boldsymbol{\mu } \right) \\
\text{s.t.}&\ \sum\limits_{n\in \mathcal{N}}{{{x}_{nk}}}=1,\ \forall k\in \mathcal{K}, \\
 &{{x}_{nk}}\in \left\{ 0,1 \right\},\ \forall n\in \mathcal{N},\forall k\in \mathcal{K}, \\
\end{split} \right.
\end{equation}
and the dual problem of \eqref{eq20} can be written as
\begin{equation}\label{eq23}
\underset{\boldsymbol{\mu }}{\mathop{\min }}\,I\left( \boldsymbol{\mu } \right).
\end{equation}
\par
Evidently, the problem \eqref{eq23} is not in a coupling form, and thus the optimum $\boldsymbol{x}$ and $\boldsymbol{y}$ can be separately achieved. According the rule of dual decomposition method, the problem \eqref{eq23} can be decomposed into
\begin{equation}\label{eq24}
I_{1}\left( \boldsymbol{\mu } \right)=\left\{ \begin{split}
  \underset{\boldsymbol{x}}{\mathop{\max }}\,& \sum\limits_{n\in \mathcal{N}}{\sum\limits_{k\in \mathcal{K}}{{{x}_{nk}}\left( {{h}_{nk}}-{{\mu }_{n}} \right)}} \\
\text{s.t.}&\sum\limits_{n\in \mathcal{N}}{{{x}_{nk}}}=1,\ \ \forall k\in \mathcal{K}, \\
 &{{x}_{nk}}\in \left\{ 0,1 \right\},\ \ \forall n\in \mathcal{N},\forall k\in \mathcal{K}, \\
\end{split} \right.
\end{equation}
\par
\noindent
and
\begin{equation}\label{eq25}
I_{2}\left( \boldsymbol{\mu } \right)=\underset{{\boldsymbol{y}}}{\mathop{\max }}\,\ \sum\limits_{n\in \mathcal{N}}{{{y}_{n}}\left\{ {{\mu }_{n}}-\log {{y}_{n}} \right\}}.
\end{equation}
\par
Now, we let ${{x}_{nk}}\left( \boldsymbol{\mu } \right)$ and ${{y}_{n}}\left( \boldsymbol{\mu } \right)$ be the maximizers of subproblems \eqref{eq24} and \eqref{eq25} respectively. Once the dual optimal ${{\boldsymbol{\mu} }^{*}}$ is found, the primal optimal solutions will be easily achieved by separately solving subproblems \eqref{eq24} and \eqref{eq25}. In this operation, any coordination doesn't exist among BSs.
\par
To search the dual optimal solution of the dual problem \eqref{eq23}, the gradient projection method \cite{39} is a good option. Specially, the Lagrange multiplier is iteratively updated by the direction of negative gradient of the objective function in the dual problem \eqref{eq23}, where the detailed direction is given by the gradient $-\nabla I\left( \boldsymbol{\mu}  \right)$. To achieve this gradient, we need to find the primal optimal solutions. That requires us to solve the subproblems \eqref{eq24} and \eqref{eq25}, which can be tackled in a distributed manner.
\par
Although the problem \eqref{eq24} is in a mixed-integer form, we can easily find the optimal association indicator ${{\boldsymbol{x}}^{*}}$, which is given by
\begin{equation}\label{eq26}
x_{nk}^{*}=\left\{ \begin{split}
  & 1,n=\arg \underset{n\in \mathcal{N}}{\mathop{\max }}\,\left\{ {{h}_{nk}}-{{\mu }_{n}} \right\},k\in \mathcal{K}, \\
 & 0,\text{otherwise}, \\
\end{split} \right.
\end{equation}
\par
The rule \eqref{eq26} shows that any user $k$ selects some BS $n$ to maximize the obtained utility ${{h}_{nk}}-{{\mu }_{n}}$. Evidently, this operation is completed by users, and thus belongs to the algorithm on user's side. The detailed descriptions can be found in user's algorithm of Algorithm 3, where $t$ represents the iteration index and ${{T}_{2}}$ is the maximal number of iterations.
\par
Note that the problem \eqref{eq25} has a simply form and can be easily solved by employing the first-order optimality condition of $I_{2}\left( \boldsymbol{\mu } \right)$ with respect to ${{y}_{n}}$. Thus, we have
\begin{equation}\label{eq27}
{{y}_{n}^{t+1}}=\exp \left( {{\mu }_{n}^{t}}-1 \right).
\end{equation}
\par
When we know the optimal association indicators and loads at t-th iteration, the Lagrange multiplier ${{\mu }_{n}}$ of any BS $n$ can be updated by
\begin{equation}\label{eq28}
\mu _{n}^{t+1}=\mu _{n}^{t}-\xi \frac{\partial I}{\partial {{\mu }_{n}}}=\mu _{n}^{t}-\xi \left( y_{n}^{t}-\sum\limits_{k\in \mathcal{K}}{x_{nk}^{t}} \right),
\end{equation}
where $\xi$ represents a sufficiently small fixed stepsize for updating ${{\mu }_{n}}$.
\par
To speed up the convergence rate of the distributed algorithm, authors in \cite{21,27} adopt Bertsekas's stepsize rule. Moreover, authors in \cite{26} propose a dual coordinate method to find optimal ${{\mu }_{n}}$. Although these methods may increase the convergence rate, the computation complexities of their algorithms are relatively high and may be unfavourable for distributed application in real system. In addition, the optimal parameter design in the former may not achieve because of too many parameters that need to be initialized. For the simplicity of implement, we just consider a fixed stepsize. Evidently, when the initial value of multiplier is properly set, the proposed algorithm can have a relatively high convergence rate. In this paper, we let the initial value of multiplier be $\log K$. Although we cannot give some proper explanations for this setting, the Monte-Carlo simulation reveals its effectiveness. As for other methods mentioned \cite{21,26,27}, we can easily employ it into the proposed algorithm, and thus we will no longer take account of them.
\par
In addition, it is easy to find that the calculation of optimal $\boldsymbol{y}$ and the update of multiplier $\boldsymbol{\mu}$ happens on BS's side, and these operations constitute an algorithm on BS's side. The detailed descriptions can be found in BS's algorithm of Algorithm 3.
\begin{table}[h]
\centering
%\caption{Algorithm 3: EEAUF}
\begin{tabular}{ll}
\toprule[1pt]
\textbf{Algorithm 3} EEAUF\\ \midrule[0.5pt]
1: \textbf{Initialization:} $t=0$, $\boldsymbol{\mu}^{t}$, $\xi $ and ${{T}_{2}}$. \\
2: \textbf{Repeat(Main Loop)}\\
3:\ \ \ \ Set $\boldsymbol{x}=\boldsymbol{0}$. \\
4:\ \ \ \ \textbf{For} $k\in \mathcal{K}$ \\
5:\ \ \ \ \ \ \ \ \textbf{\%Algorithm at User Terminal $k$\%} \\
6:\ \ \ \ \ \ \ \ User $k$ selects BS ${{n}^{*}}$ that satisfies the following condition:\\
7:\ \ \ \ \ \ \ \ \ \ \ \ \ ${{n}^{*}}=\arg \underset{n\in \mathcal{N}}{\mathop{\max }}\,\left\{ {{h}_{nk}}-{{\mu }_{n}} \right\}$.\\
8:\ \ \ \ \ \ \ \ Set ${{x}_{{{n}^{*}}k}}=1$.\\
9:\ \ \ \ \textbf{EndFor} \\
10:\ \ \ \ \textbf{For} $n\in \mathcal{N}$ \\
11:\ \ \ \ \ \ \ \ \textbf{\%Algorithm at Base Station $n$\%} \\
12:\ \ \ \ \ \ \ \ ${{y}_{n}^{t+1}}=\exp \left( {{\mu }_{n}^{t}}-1 \right)$.\\
13:\ \ \ \ \ \ \ \ $\mu _{n}^{t+1}=\mu _{n}^{t}-\xi \left( y_{n}^{t}-\sum\limits_{k\in \mathcal{K}}{x_{nk}^{t}} \right)$.\\
14:\ \ \ \ \textbf{EndFor} \\
15:\ \ \ \ $t=t+1$ \\
16:\textbf{Until} convergence or $t={{T}_{2}}$\\ \bottomrule[0.5pt]
\end{tabular}
\label{tab1}
\end{table}
\par
It is easy to find that the formula \eqref{eq28} has some special meanings. This formula shows that the update of any multiplier $\mu_{n}$ obeys the supply and demand rule. In this rule, $\sum\nolimits_{k\in \mathcal{K}}{{{x}_{nk}}}$ represents serving demand of any BS $n$, ${{y}_{n}}$ is its available service and the multiplier $\mu_{n}$ denotes its service cost. This cost will go up if the demand $\sum\nolimits_{k\in \mathcal{K}}{{{x}_{nk}}}$ exceeds the supply ${{y}_{n}}$ and vice versa. From the perspective of user association, some overloaded BSs may increase their costs to prevent users accessing, but some underloaded BSs may decrease their costs to attract more users.
\par
During user association, any BS $n$ needs to broadcast the message $\mu_{n}$ to all users and any user $k$ just feedbacks the association indicator ${{x}_{{{n}^{*}}k}}=1$ to the expected BS ${{n}^{*}}$. To this end, the whole association procedure should interactively perform the algorithms on user's and BS's side. The convergence of distributed algorithm can be proven by employing results of subgradient method, which is stated in the following lemma \cite{40}.
\\
\par
\noindent
\textbf{Lemma 3.} The multiplier $\boldsymbol{\mu}$ is updated by utilizing the rule \eqref{eq28}, then
\begin{equation}\label{eq29}
I_{best}^{t}-{{I}^{*}}\le \frac{\left\| {{\boldsymbol{\mu}}^{1}}-{{\boldsymbol{\mu}}^{*}} \right\|_{2}^{2}+t{{\xi }^{2}}{{\tau }^{2}}}{2t\xi },
\end{equation}
where ${{\boldsymbol{\mu}}^{*}}$ is a point that minimizes $I$; ${{I}^{*}}=I\left( {{\boldsymbol{\mu}}^{*}} \right)$; $I_{best}^{t}=\underset{i=1,\cdots ,t}{\mathop{\min }}\,I\left( {{\boldsymbol{\mu}}^{i}} \right)$; ${\tau }$ is an upper bound of subgradient $\nabla I\left( \boldsymbol{\mu } \right)$. The equation \eqref{eq29} shows that the subgradient method converges to within some range of the optimal value. Specially, the subgradient method can achieve an $\varepsilon$-suboptimal point within a finite number of iterations, where $\varepsilon ={\left( \left\| {{\boldsymbol{\mu }}^{1}}-{{\boldsymbol{\mu }}^{*}} \right\|_{2}^{2}+t{{\xi }^{2}}{{\tau }^{2}} \right)}/{2t\xi }\;$.
\par
Furthermore, there also exist some results for the equation \eqref{eq29}, which are given in the following lemma \cite{40}.
\\
\par
\noindent
\textbf{Lemma 4.} The righthand side of \eqref{eq29} converges to ${\xi {{\tau }^{2}}}/{2}$ as $t\to \infty$. Thus, when $\xi$ is given, $I_{best}^{t}$ converges to within ${\xi {{\tau }^{2}}}/{2}\text{-suboptimal}$. Moreover, $I_{best}^{t}-{{I}^{*}}\le \xi {{\tau }^{2}}$ within at most ${\left\| {{\boldsymbol{\mu}}^{1}}-{{\boldsymbol{\mu}}^{*}} \right\|_{2}^{2}}/{{{\xi }^{2}}{{\tau }^{2}}}$ iterations.
\\
\par
Now, we give the complexity analyses of proposed algorithms. Similar to Algorithm 1, the Algorithm 3 also belongs to the iterative algorithm. Obviously, the former has a computation complexity of $\mathcal{O}\left( {{T}_{1}}NK \right)$, and the latter has a computation complexity of $\mathcal{O}\left( {{T}_{2}}NK \right)$. Unlike other algorithms, the Algorithm 2 doesn't belong to the iterative algorithm and has a simply form, and its computation complexity is $\mathcal{O}\left( NK \right)$. Moreover, during user association, Algorithm 2 doesn't require any exchanged information between users and BSs. However, the amount of exchanged information in Algorithm 1 is proportional to $NK$, and the one of Algorithm 3 is $N+K$.
\section{Numerical Results}
We assume that the inter-site distance between any two MBSs is 1000 m, the distance between any two PBSs is larger than 40 m, the distance between PBS and MBS is larger than 75 m, the distance between user and MBS is larger than 35 m, the distance between user and PBS is larger than 10 m \cite{41}. We assume that the maximal transmit power of each user is 23 dBm \cite{24}, the circuit power is 0.1 W \cite{42}, the noise power spectral density is -174 dBm/Hz and the processing gain $\kappa $ is 128 \cite{28}. For the propagation environment, we adopt the pathloss model \cite{43} ${{l}_{nk}}=128.1+37.6\log 10\left( {{d}_{nk}} \right)$ dB and ${{l}_{nk}}=140.7+36.7\log 10\left( {{d}_{nk}} \right)$ dB for MBS and PBS respectively, where ${{d}_{nk}}$ is the distance between the user $k$ and BS $n$ in kilometers. Meanwhile, we consider the log-normal shadowing with a standard deviation 8 dB.
\par
In the simulation, we will investigate three types of energy-efficient association schemes under open loop power control. Meanwhile, we introduce other association schemes for comparison, which include strategies MARA and AUF. The former performs user association without considering the cell load, but the latter mentions it during association. In terms of association performance, we mainly focus on the average throughput (rate), the average energy efficiency, the whole energy efficiency, the load balancing level and the ratio of supported users scattered into networks. To investigate these performance indices, we take two types of network parameters into account, which include the density (number) of users in each macrocell and the one of PBSs in each macrocell.
\par
\begin{figure}[!t]
\centering
\centerline{\includegraphics[width=4in]{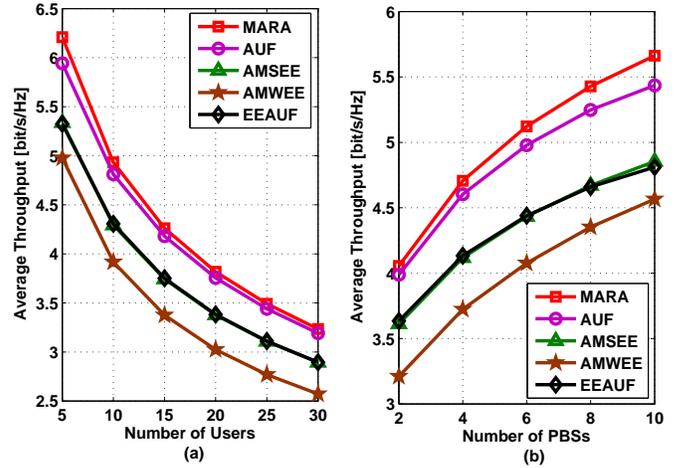}}
\caption{The average throughput comparison: (a) the impact of the number of users; (b) the impact of the number of PBSs.}
\label{fig2}
\end{figure}
Fig. \ref{fig2} shows the average throughput of strategies MARA, AUF, AMSEE, AMWEE and EEAUF, where Fig. \ref{fig2} (a) investigates the influence of the number of users on the average throughput and Fig. \ref{fig2} (b) shows the impact of the number of PBSs on this performance index. Note that the number of users refers to the one of users scattered into each macrocell, and the number of PBSs also refers to the one of PBSs scattered into each macrocell. According the definition of achievable rate, we know that the throughput should decrease with the number of users due to severer and severer interference, which is illustrated in Fig. \ref{fig2} (a). When the number of PBSs increases, the distance between users and PBSs becomes shorter. Thus, the achievable rate will increases with number of PBSs, which means the average throughput will increases with this number in Fig. \ref{fig2} (b). As illustrated in Fig. \ref{fig2}, the strategy MARA achieves the highest throughput among all strategies since it just maximizes the sum rate. Unlike strategy MARA, the strategy AUF also needs to consider the cell load for load balancing during user association, which means that not all users can always select some BSs with high achievable rates. Thus, the strategy AUF has a lower throughput than the strategy MARA. In addition, Fig. 2 also reveals that the strategy AMWEE achieves the lowest throughput, and strategies AMSEE and EEAUF obtain almost the same throughput.
\par
\begin{figure}[!t]
\centering
\centerline{\includegraphics[width=4in]{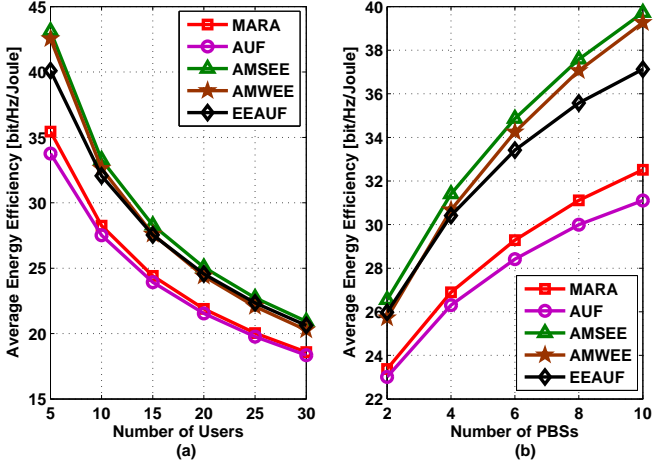}}
\caption{The average energy efficiency comparison: (a) the impact of the number of users; (b) the impact of the number of PBSs.}
\label{fig3}
\end{figure}
Fig. \ref{fig3} shows the average energy efficiencies of strategies MARA, AUF, AMSEE, AMWEE and EEAUF, where Fig. \ref{fig3} (a) investigates the influence of the number of users on the average energy efficiency and Fig. \ref{fig3} (b) shows the impact of the number of PBSs on this performance index. Since the achievable rate decreases with the number of users and the transmit power is not unaffected by this number, the average energy efficiency should decrease with the number of users in Fig. \ref{fig3} (a). As mentioned in Fig. \ref{fig3}, the distance between users and PBSs becomes shorter when the number of PBSs increases. Thus, the achievable rate will increases with number of PBSs, and the transmit power of users may decrease. Consequently, the average throughput will increase with this number in Fig. \ref{fig3} (a). In the strategy AMSEE, all users try to maximize their own energy efficiencies during association, and this strategy achieves the highest average energy efficiency among all strategies. As shown in Fig. \ref{fig3}, other energy-efficient strategies achieve lightly lower average energy efficiency than the strategy AMSEE, and the strategy AMWEE obtains lightly higher average energy efficiency than the strategy EEAUF. Since the strategy MARA has a higher average energy efficiency than the strategy AUF, the former may provide higher average energy efficiency than the latter.
\par
\begin{figure}[!t]
\centering
\centerline{\includegraphics[width=4in]{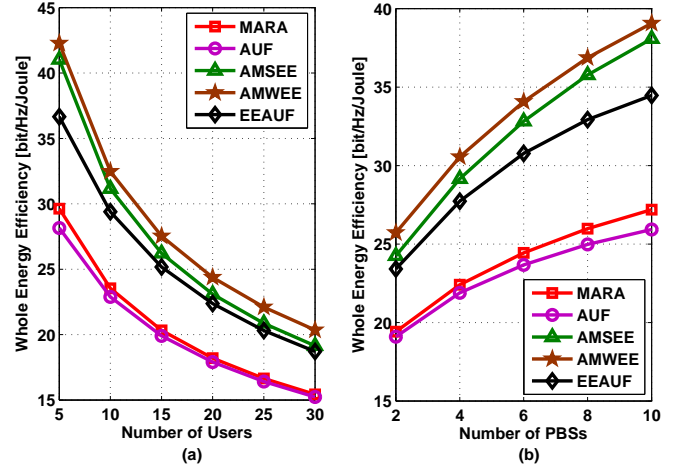}}
\caption{The whole energy efficiency comparison: (a) the impact of the number of users; (b) the impact of the number of PBSs.}
\label{fig4}
\end{figure}
Fig. \ref{fig4} shows the whole energy efficiencies of strategies MARA, AUF, AMSEE, AMWEE and EEAUF, where Fig. \ref{fig4} (a) investigates the influence of the number of users on the whole energy efficiency and Fig. \ref{fig4} (b) shows the impact of the number of PBSs on this performance index. Similar to the Fig. \ref{fig4}, the whole energy efficiency should decrease with the number of users in Fig. \ref{fig4} (a) and it should increase with the number of PBSs in Fig. \ref{fig4} (b). Since the aim of strategy AMWEE is to maximize the whole energy efficiency, and this strategy achieves the highest whole energy efficiency among all strategies. As shown in Fig. \ref{fig4}, other energy-efficient strategies achieve lightly lower whole energy efficiency than the strategy AMWEE, and the strategy AMSEE obtains lightly higher whole energy efficiency than the strategy EEAUF. Similarly, the strategy MARA may provide higher whole energy efficiency than the strategy AUF since the former has a higher average throughput (rate) than the latter.
\par
Through direct observation, we can easily find that the strategy EEAUF achieves the lowest average and whole energy efficiency. The reason for this is probably that the offloading (load balancing) operation reduce the achievable rates of associated users. This impact also exists in strategies MARA and AUF.
\par
To measure the load balancing level of the whole network, we introduce Jain's fairness index \cite{44} as follows.
\begin{equation}\label{eq30}
J=\frac{{{\left( \sum\nolimits_{n\in \mathcal{N}}{{{y}_{n}}} \right)}^{2}}}{N\sum\nolimits_{n\in \mathcal{N}}{y_{n}^{2}}},
\end{equation}
where $\sum\nolimits_{k\in \mathcal{K}}{{{x}_{nk}}}={{y}_{n}}$ is the load of BS ${n}$. A larger $J$, taken value from the interval $\left[ \frac{1}{N},1 \right]$,  means a more balanced load distribution.
\par
\begin{figure}[!t]
\centering
\centerline{\includegraphics[width=4in]{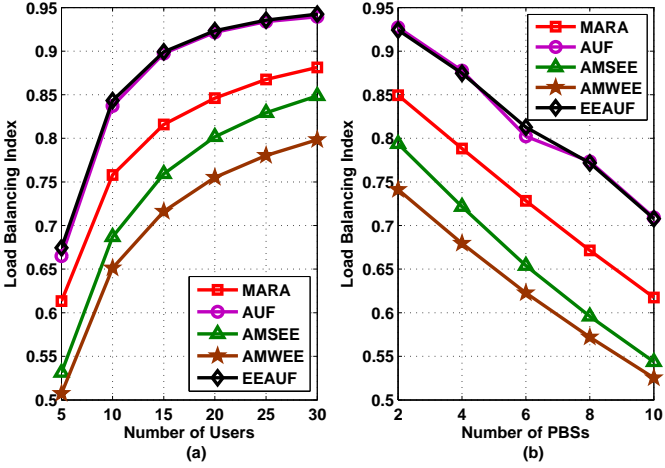}}
\caption{The load balancing level comparison: (a) the impact of the number of users; (b) the impact of the number of PBSs.}
\label{fig5}
\end{figure}
Fig. \ref{fig5} shows the load balancing levels of strategies MARA, AUF, AMSEE, AMWEE and EEAUF, where Fig. \ref{fig5} (a) investigates the influence of the number of users on load balancing level and Fig. \ref{fig5} (b) shows the impact of the number of PBSs on this performance index. According the simulation model, we know that the user distribution gets even when the number of users increases. Thus, the load distribution has more opportunities to achieve a higher balancing level, which means that the load balancing level increases with the number of users in Fig. \ref{fig5} (a). Unlike Fig. \ref{fig5} (a), the load balancing level should decrease with the number of PBSs in Fig. \ref{fig5} (b). According to the simulation parameters, it is easy to find that the value range of PBS's pathloss often contains the one of MBS's pathloss. Thus, users in the signal strength-based association strategy have more opportunities to select PBSs. As we know, all strategies depend on the achievable rate. That means more PBSs often attract more users for themselves. In Fig. \ref{fig5}, two types of load balancing strategies, including AUF and EEAUF, achieve the highest load balancing level among all strategies. In addition, the load balancing level of strategy EEAUF may be slightly higher than the strategy AUF due to the impact of logarithmic function on transmit power, i.e., fairness of transmit power. Since the energy-efficient strategies enhance the impact of pathloss, the strategies AMSEE and AMWEE achieve lower load balancing level than strategy MARA. Moreover, Fig. \ref{fig5} also reveals that the strategy AMSEE has slightly higher load balancing level than the strategy AMWEE.
\par
\begin{figure}[!t]
\centering
\centerline{\includegraphics[width=4in]{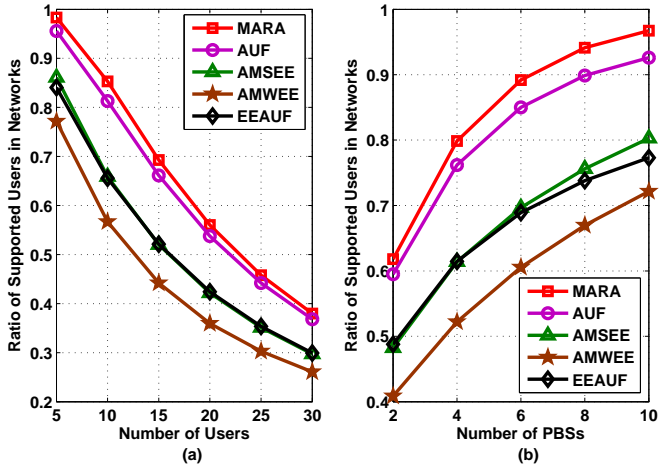}}
\caption{The supported ratio comparison: (a) the impact of the number of users; (b) the impact of the number of PBSs.}
\label{fig6}
\end{figure}
Fig. \ref{fig6} shows the supported ratio of strategies MARA, AUF, AMSEE, AMWEE and EEAUF, where Fig. \ref{fig6} (a) investigates the impact of the number of users on the supported ratio and Fig. \ref{fig6} (b) shows the impact of the number of PBSs on this performance index. The supported ratio refers to the ratio of special users to all users in networks, and these special users have the achievable rates that are larger than target rates. Since a large (achievable) rate often means a high supported ratio, the Fig. \ref{fig6} should have the same trend with Fig. \ref{fig2}.
\par
Although we cannot give the proper explanations for the performance trends of some association strategies sometimes, simulation results show that the designed energy-efficient association strategies can always optimize their own performance metrics, which means the effectiveness of designed strategies can be guaranteed. In other words, the energy-efficient strategies can maximize the corresponding performance indices in simulation figures.
\par
\begin{figure}[!t]
\centering
\centerline{\includegraphics[width=4in]{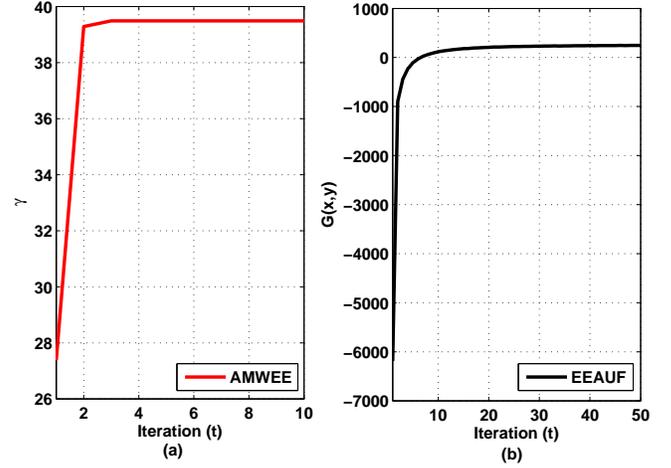}}
\caption{The convergence of proposed algorithms: (a) Algorithm 1 (AMWEE); (b) Algorithm 3 (EEAUF).}
\label{fig7}
\end{figure}
Fig. \ref{fig7} (a) and Fig. \ref{fig7} (b) show the convergences of Algorithm 1 (AMWEE) and Algorithm 2 (EEAUF) respectively, where $t$ is iteration index. Fig. \ref{fig7} shows that the proposed algorithms have fast convergence rates and can be well implemented in real system.
\section{Conclusion}
In this paper, we propose three types of energy-efficient association schemes under open loop power control for uplink HCNs, and then design effective algorithms for these schemes. At last, we give the convergence proofs of proposed algorithms and provide the complexity analyses. To highlight the effectiveness of proposed schemes, we introduce other association schemes for comparison. In the numerical simulation, we investigate the influences of different network parameters on the association performance. Simulation results show that all energy-efficient association schemes achieve their own (optimization) targets.

\section*{Acknowledgment}
This work was supported by the National Natural Science Foundation of China under Grants 61372101, 61422105 and 61271018, the National Science and Technology Major Project of China under Grants 2013ZX03003006-002 and 2012ZX03004-005-003, the Natural Science Foundation of Jiangsu Province under Grant BK20130019, the Research Project of Jiangsu Province under Grant BE2012167, the Program for New Century Excellent Talents in University under Grant NCET-11-0088,
the Joint Funds of the National Natural Science Foundation of China under Grant U1404615, and the Open Funds of State Key Laboratory of Millimeter Waves under Grant K201504.

\end{document}